\documentclass{aastex631}

\usepackage{bm}
\usepackage{graphicx}
\usepackage{hyperref}
\usepackage{natbib}
\usepackage[figuresright]{rotating}
\usepackage{subfigure}
\usepackage{makecell}
\begin{document}

\title{Pulsed Iron line Emission from the First Galactic Ultraluminous X-ray Pulsar Swift J0243.6+6124}

\author{Y.X. Xiao}
\affiliation{Key Laboratory of Particle Astrophysics, Institute of High Energy Physics, Chinese Academy of Sciences, Beijing 100049, China}

\author{Y.J. Xu$^{\star}$}
\affiliation{Key Laboratory of Particle Astrophysics, Institute of High Energy Physics, Chinese Academy of Sciences, Beijing 100049, China}

\author{M.Y. Ge$^{\star}$}
\affiliation{Key Laboratory of Particle Astrophysics, Institute of High Energy Physics, Chinese Academy of Sciences, Beijing 100049, China}

\author{F.J. Lu}
\affiliation{Key Laboratory of Particle Astrophysics, Institute of High Energy Physics, Chinese Academy of Sciences, Beijing 100049, China}
\affiliation{Key Laboratory of Stellar and Interstellar Physics and School of Physics and Optoelectronics, Xiangtan University, Xiangtan 411105, Hunan, China}

\author{S.N. Zhang}
\affiliation{Key Laboratory of Particle Astrophysics, Institute of High Energy Physics, Chinese Academy of Sciences, Beijing 100049, China}
\affiliation{University of Chinese Academy of Sciences, Chinese Academy of Sciences, Beijing 100049, China}

\author{S. Zhang}
\affiliation{Key Laboratory of Particle Astrophysics, Institute of High Energy Physics, Chinese Academy of Sciences, Beijing 100049, China}

\author{L. Tao}
\affiliation{Key Laboratory of Particle Astrophysics, Institute of High Energy Physics, Chinese Academy of Sciences, Beijing 100049, China}

\author{J.L. Qu}
\affiliation{Key Laboratory of Particle Astrophysics, Institute of High Energy Physics, Chinese Academy of Sciences, Beijing 100049, China}

\author{P.J. Wang}
\affiliation{Key Laboratory of Particle Astrophysics, Institute of High Energy Physics, Chinese Academy of Sciences, Beijing 100049, China}

\author{L.D. Kong}
\affiliation{Institut f{\"u}r Astronomie und Astrophysik, Sand 1, 72076 T{\"u}bingen, Germany}

\author{Y.L. Tuo}
\affiliation{Institut f{\"u}r Astronomie und Astrophysik, Sand 1, 72076 T{\"u}bingen, Germany}

\author{Y. You}
\affiliation{Key Laboratory of Particle Astrophysics, Institute of High Energy Physics, Chinese Academy of Sciences, Beijing 100049, China}

\author{S.J. Zhao}
\affiliation{Key Laboratory of Particle Astrophysics, Institute of High Energy Physics, Chinese Academy of Sciences, Beijing 100049, China}

\author{J.Q. Peng}
\affiliation{Key Laboratory of Particle Astrophysics, Institute of High Energy Physics, Chinese Academy of Sciences, Beijing 100049, China}

\author{Y.F. Du}
\affiliation{Key Laboratory of Particle Astrophysics, Institute of High Energy Physics, Chinese Academy of Sciences, Beijing 100049, China}

\author{Y.H. Zhang}
\affiliation{Key Laboratory of Particle Astrophysics, Institute of High Energy Physics, Chinese Academy of Sciences, Beijing 100049, China}

\author{W.T. Ye}
\affiliation{Key Laboratory of Particle Astrophysics, Institute of High Energy Physics, Chinese Academy of Sciences, Beijing 100049, China}

\begin{abstract}

We report the phase-resolved spectral results of the first Galactic Pulsating Ultra-Luminous X-ray source (PULX) Swift J0243.6+6124, modeling at its 2017-2018 outburst peak using data collected by the Hard X-ray Modulation Telescope (\textit{Insight}-HXMT). The broad energy coverage of \textit{Insight}-HXMT allows us to obtain more accurate spectral continuum to reduce the coupling of broad iron line profiles with other components. We use three different continuum spectrum models but obtain similar iron line results. For the first time, we detected the pulse characteristics of the broad iron line in a PULX. The variation in width and intensity of this iron line with $\sigma \sim 1.2-1.5$\,keV has a phase offset of about 0.25 from the pulse phase. We suggest that the uneven irradiation of the thick inner disk by the accretion column produces the modulated variation of the broad iron line. In addition, the non-pulsed narrow line is suggested to come from the outer disk region.

\end{abstract}

\keywords{pulsars: individual (Swift J0243.6+6124), X-rays: binaries, accretion: accretion pulsar}

\section{Introduction} 

Super-Eddington accreting pulsars can power ultraluminous X-ray sources (ULX), which are among the brightest astrophysical sources (above $\sim$ $10^{39}$ erg s$^{-1}$) in the X-ray sky (for a review see Kaaret et al. \citeyear{2017ARA&A..55..303K}). Since \cite{2014Natur.514..202B} first observed X-ray pulses from an ULX (M82 X-2), many pulsating ULX (PULX) sources have been discovered one after another \citep{2016ApJ...831L..14F,2017MNRAS.466L..48I,2017Sci...355..817I,2018MNRAS.476L..45C,2020MNRAS.495.2664C}. A pulsar in an X-ray binary system is able to produce such a high luminosity due to the accretion geometry created by their compact stellar surfaces and complex magnetic fields: an accretion column can be formed near the magnetic poles of the pulsar, and strong radiation escaped from the sides of the column will produce a fan-beam emission. The accretion column illuminates a large surrounding area, allowing us to study the spatial distribution of the surrounding matter. 

Many astronomical X-ray sources produce fluorescent iron lines by reprocessing of illumination from the central engine by surrounding materials such as the accretion disk. The emission lines can be skewed and broadened by the combined effects from Doppler effect, gravitational redshift, and Compton Scattering. Therefore, it is possible for us to analyze the ionization degree, kinematics, and geometric distribution of materials around the radiation area through the spectral structure shown by the iron lines, especially in neutron star systems usually with complex accretion structures. In fact, pulsed iron line emissions associated with rotation are detected in some high mass X-ray binaries (HXMBs) such as Cen X-3 \citep{1993ApJ...408..656D}, Her X-1 \citep{1994ApJ...437..449C}, LMC X-4 \citep{2017AstL...43..175S}, 4U 1538-522 \citep{2014ApJ...792...14H}, GX 301-2 \citep{2018MNRAS.480.4746L}, and V 0332+53 \citep{2021MNRAS.506.2156B}, but none of them are PULX. The known PULXs are distant ($>$ 80\,kpc, except our target), and the limited number of the pulsed photons is difficult to support the analysis of the emission line. We need higher statistics to resolve the behavior of the iron line in order to probe the geometry at the super-Eddington accretion state of a PULX.

The transient X-ray source Swift J0243.6+6124 was discovered by the \textit{Neil Gehrels Swift Observatory} in October 2017, when it reached a flux of $\sim$ 80\,mCrab at the rising phase of a giant outburst \citep{2017GCN.21960....1C,2017ATel10809....1K}. The detection of the pulse period of $\sim$ 9.86\,s and the optical counterpart of a Be star in this source, identified it as a Be X-ray binary pulsar \citep{2017ATel10809....1K,2017ATel10907....1G,2017ATel10812....1J,2017ATel10822....1K,2020A&A...640A..35R}. The distance to this Be star measured by Gaia DR2 parallax is 6.8\,kpc \citep{2018AJ....156...58B}. Together with the peak flux exceeding 8 Crab, this distantce reveals that the source is the first Galactic PULX. Such a bright source makes a good laboratory for studying the physics of super-Eddington accretion.

In the super-Eddington state of Swift J0243.6+6124, the broad iron line complex has been resolved into components centered at 6.4\,keV, 6.67\,keV, and 6.98\,keV by \cite{2019ApJ...885...18J} using high energy resolution data from \textit{NICER} and \textit{NuSTAR}. The putative neutral 6.4\,keV line has a fairly large width, with $\sigma\sim 1.67$\,keV when approaching the outburst peak. The edge feature at 7.1\,keV is obvious at this time. \cite{2020ApJ...902...18K} reported the evolution of the spectral parameters during the outburst with \textit{Insight}-HXMT observations, where $\sigma$ reached 1.7\,keV at the peak. They also gave phase-resolved spectral parameters at the outburst peak \citep{2022ApJ...933L...3K}, but did not analyze the properties of the line emission in details. \cite{2022MNRAS.516.1601B} also studied the phase-resolved spectrum using the reflection model with \textit{NuSTAR} observations, and obtained a pulsed reflection fraction in the super Eddington state. We note that such a broad iron line is difficult to be separated from the continuum component in the spectrum if the energy range is not wide enough, as the continuum model and cutoff power-law plus blackbody for accretion pulsars may be coupled to the broad line. Therefore, we hope that more accurate phase-resolved line parameters can be obtained by conducting detailed spectral analysis of the broad-band \textit{Insight}-HXMT observations.
In section \ref{obs}, we list the observations and methods of data reduction. In section \ref{results}, we present the spectral results for different models. We discuss our results in Section \ref{discuss} and make conclusions in Section \ref{summary}. 

\section{Observations and Data Reduction} \label{obs}

\textit{Insight}-HXMT carries three slat-collimated X-ray telescope: the High energy X-ray telescope (HE),  the Medium Energy X-ray telescope (ME), and the Low Energy X-ray telescope (LE). Its broad energy band (1$-$250\,keV), large detection area (5100 cm$^2$ in 20$-$250\,keV for HE) and the small dead time make it a powerful satellite in X-ray spectral studies of bright X-ray sources \citep{2019SCPMA..6229502Z,2020SCPMA..6349502Z}.

 {\it Insight}-HXMT made 102 pointing observations of Swift J0243.6+6124 during its 2017$-$2018 outburst. The \textit{Insight}-HXMT Data Analysis Software (\texttt{HXMTDAS}) v2.04 with default filters is used to reduce the data. These include screening data with elevation angle (ELV) $>$ 10$^{\circ}$, geometric cutoff rigidity (COR) $>$ 8\,GeV, offset for the point position $\leq$ 0.04$^{\circ}$, and time beyond 300\,s to the South Atlantic Anomaly (SAA). We use \texttt{HXMTDAS} tasks \texttt{hespecgen}, \texttt{mespecgen} and \texttt{lespecgen} to generate the spectra. The background spectra are estimated with \texttt{hebkgmap}, \texttt{mebkgmap} and \texttt{lebkgmap}. The backgrounds are estimated via version 2.0.9 of the current standard \textit{Insight}-HXMT background model \citep{2020JHEAp..27...24L,2020JHEAp..27...14L,2020JHEAp..27...44G}. And the response matrices are created by \texttt{herspgen}, \texttt{merspgen} and \texttt{lerspgen} tasks.

We select ten exposures (Table \ref{table:10 observations}) near the brightest epoch from 58060 to 58066 (MJD) as used by \cite{2022ApJ...933L...3K}, because the spectral parameters remain stable for these ten exposures \citep{2020ApJ...902...18K}. The photon arriving times are corrected by solar system barycenter and binary-orbiting modulation (refer to the parameters in Table \ref{table:orbit}). The parameters of binary orbit are taken from the website of Fermi/GBM Accreting Pulsar Histories\footnote{https://gammaray.msfc.nasa.gov/gbm/science/pulsars/lightcurves/swiftj0243.html}. The daily spin frequencies taken by the Fermi/GBM allows us to use cubic spline interpolation to fit the frequency with time $\nu(t)$. Considering the changing pulse-period, the phase-coherent pulse profiles can be derived by calculated a sequential pulse phase $\phi(t)$ for corrected event time $t$ as
\begin{equation}
    \phi(t)=\int_{t_{0}}^{t} \nu(\tau)d\tau+\phi_{0} \, ,\\
\end{equation}
where $t_{0}$ is fixed at 58027.499066\,(MJD), which is the epoch of the first Fermi/GBM periodicity detection \citep{2020ApJ...896..124S}. We set $\phi_{0}=0.07$ to make the minimum region of pulse profile fall within the phase 0.1$-$0.2. To generate the phase-resolved spectrum, we use \texttt{hxmtscreen} to further filter the photons within a specific phase interval, and then generate the spectrum following procedure mentioned before. Finally, we combine the spectra, the background and the response corresponding to the same phase interval. We also combine the spectra of these ten observations as the phase-averaged spectrum.

\begin{table}
\small
\caption{Information of 10 \textit{Insight}-HXMT observations used in this work}
\label{table:10 observations}
\medskip
\begin{center}
\begin{tabular}{l l c c}
\hline \hline
ObsID  &  Time  &  LE Exposure  &  1$-$250\,keV flux\\
  &  (MJD)  &  (s)  &  ($10^{-7}$\, erg\,cm$^{-2}$\,s$^{-1}$)\\
\hline
P011457701701  &  58060.3  &  239.4  &  $3.23^{+0.01}_{-0.02}$\\
P011457701702  &  58060.4  &  2214  &  $3.16^{+0.01}_{-0.01}$\\
P011457701703  &  58060.6  &  2691  &  $3.18^{+0.01}_{-0.01}$\\
P011457701704  &  58060.7  &  1257  &  $3.17^{+0.01}_{-0.01}$\\
P011457701801  &  58061.3  &  299.2  &  $3.38^{+0.01}_{-0.02}$\\
P011457701901  &  58062.6  &  3739  &  $3.37^{+0.01}_{-0.01}$\\
P011457701902  &  58062.8  &  2421  &  $3.37^{+0.01}_{-0.01}$\\
P011457702001  &  58064.1  &  1252  &  $3.61^{+0.01}_{-0.01}$\\
P011457702101  &  58065.6  &  3763  &  $3.55^{+0.01}_{-0.01}$\\
P011457702102  &  58065.8  &  2436  &  $3.50^{+0.01}_{-0.01}$\\

\hline
\end{tabular}
\\
\end{center}
\end{table}

\begin{table}
\small
\caption{Coordinate and orbit parameters of Swift J0243.6+6124}
\label{table:orbit}
\medskip
\begin{center}
\begin{tabular}{l c | c}
\hline
\hline
R.A. & ($^{\circ}$) &  40.9180 \\
Decl.& ($^{\circ}$) &  61.4341 \\
\hline
$P_{\rm orbit}$ & (days) &  27.698899\\
$T_{0}$ & (MJD)  &  58102.97476560854\\
$e$ & &  0.1029\\
$ax$sin$i$ & (lt-s) &  115.531\\
$\omega$ & ($^{\circ}$) &  -74.05\\ 
\hline

\hline
\end{tabular}
\\
\end{center}
\end{table}

The spectrum is fitted by the \texttt{XSPEC} 12.11.1 software package \citep{1996ASPC..101...17A}. We use the \texttt{grppha} to group minimum photon counts for HE, ME, LE to be 200, 50, and 30, respectively. The energy bands for LE, ME and HE used for spectral fitting are 1$-$10\,keV, 10$-$30\,keV and 28$-$100\,keV, respectively. There are some calibration residual structures of Si and Ag around 1.6$-$1.9\,keV and 19$-$23\,keV, respectively, which have been ignored when making spectral fitting \citep{2020JHEAp..27...64L}. Data in these two energy regions are neglected in subsequent fittings. We further set the systematic uncertainty to 0.5\% in all energy band on the basis of \cite{2022ApJ...933L...3K}. The error of a single parameter of interest is quoted at the 90\% confidence level.

\section{Results} \label{results}

\subsection{Phase-averaged spectrum}

\begin{figure}
    \centering
    \includegraphics[width=0.6\textwidth]{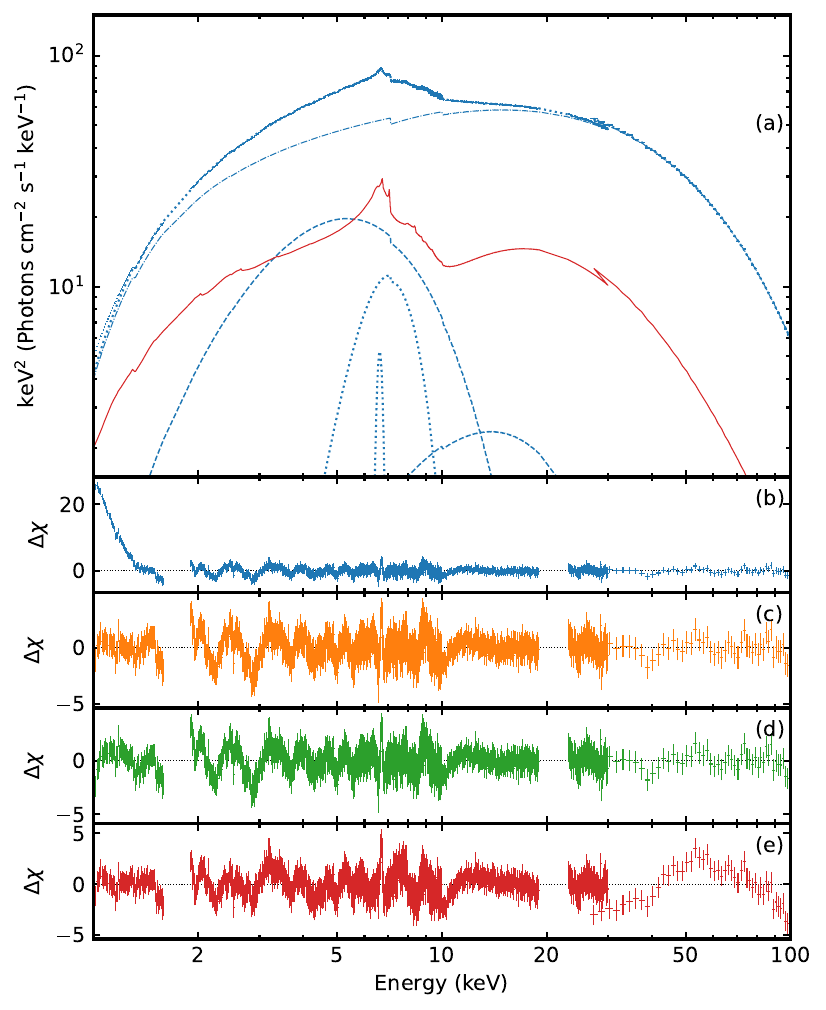}
    \caption{The phase-averaged spectrum (a) and reduce residuals fitted with Model-I (b), Model-II (c) Model-III (d) and Model-IV (e). The energy range of Model-I fitting is 2-100 keV, but the residuals is shown in 1-100 keV. The dotted lines represent the iron lines. The solid red line shows the reflection component. The two dashed lines represent the lower and higher blackbody components while the dotted dash line draws the cut-off power-law component. }
    \label{fig:average spec}
\end{figure}

The model consisting of a single blackbody plus a cutoff power-law is a good approximation for the continuum spectrum of an accretion-powered X-ray pulsar \citep{1983ApJ...270..711W,1997ApJS..113..367B}.
We first apply Model-I: $constant \times tbabs \times (bbodyrad1+bbodyrad2+gaussian1+gaussian2+cutoffpl) \times edge$ to the phase-averaged spectrum in 1$-$100\,keV. When fitting the hump structure in 10$-$30\,keV, the additional $bbodyrad2$ ($kT \sim 3.5\,\rm keV$ , $R_{\rm bb} \sim 1\,\rm km$) can reduce $\chi^{2}$ by 96 and its necessity has also been mentioned in previous works \citep{2020ApJ...902...18K,2022ApJ...933L...3K,2019ApJ...873...19T}.
The Tuebingen-Boulder model $tbabs$ \citep{2000ApJ...542..914W} is used to account for the absorption due to the interstellar medium (ISM). After fitting by Model-I, a sharp line feature at 6.7\,keV with $\sigma < 10^{-4}$\,keV and an emission-like feature around 9\,keV with $\sigma \sim 0.18\,$keV still remain (Figure \ref{fig:average spec}, b), with intensities approximately 0.02 and 0.25 times that of $gauss2$, respectively. When combining the integrated spectrum of 10 observations, the systematic error dominates the spectral residuals. Compared to other residual distributions, these structures are weak calibrated residuals \citep{2020JHEAp..27...64L}, which have little effect on the fitting results of the iron lines. We note that there is a soft excess in 1$-$2\,keV. Soft excess features are common in accreting pulsars with X-ray luminosity $L_{\rm X}\gtrsim10^{38}\, \rm erg\, s^{-1}$ \citep{2004ApJ...614..881H}. 
Therefore, we should ignored the 1$-$2\,keV energy range for Model-I. The residuals are shown by the blue dots of 2$-$100\,keV in Figure \ref{fig:average spec} (b). The $\chi^{2}$ fitted by Model-I in the energy range of 1$-$100\,keV is 1850 (1334 d.o.f) while in 2$-$100\,keV is 1371 (1252 d.o.f).

\begin{table}
\small
\caption{Model definitions in this work}
\label{table:4 models}
\medskip
\begin{center}
\begin{tabular}{l l}
\hline \hline
Model-I    &  $constant \times tbabs \times (bbodyrad1+bbodyrad2+gaussian1+gaussian2+cutoffpl) \times edge$\\
Model-II   &  $constant \times tbabs \times (bbodyrad1+bbodyrad2+bbodyrad3+gaussian1+gaussian2+cutoffpl) \times edge$\\
Model-III  &  $constant \times tbabs \times tbpcf \times (bbodyrad1+bbodyrad2+gaussian1+gaussian2+cutoffpl) \times edge$\\
Model-IV   &  $constant \times tbabs \times tbpcf \times (bbodyrad1+relxilllp+cutoffpl)$\\
\hline

\hline
\end{tabular}
\\
\end{center}
\end{table}

Careful fitting of the continuum can help to limit the parameters of broad iron line. The broad iron line complex can be fitted by double Gaussian with an absorption edge. It was reported by \cite{2019ApJ...885...18J} near the outburst peak of Swift J0243.6+6124. The broad line with $\sigma \sim 1.3\,$keV is strongly coupled with the blackbody component with $kT \sim$ 1$-$2\,keV in spectrum fitting. Removing the blackbody model causes the width of the broad line to increase to more than 2\,keV. By taking into account of the energy range of 1$-$2\,keV, we can thus obtain more accurate blackbody parameter measurements and thus constraining the broad iron line better. For the soft excess in 1$-$2\,keV, we draw on the work of \cite{2019ApJ...873...19T} to add a blackbody component with a $kT \sim$ 0.1\,keV for Model-II: $constant \times tbabs \times (bbodyrad1+bbodyrad2+bbodyrad3+gaussian1+gaussian2+cutoffpl) \times edge$. The fitted residuals are shown by the orange dots in Figure \ref{fig:average spec} (c). Adding this model resulted in a reduction of 379 in $\chi^{2}$ with 1332 d.o.f.. If 1$-$2\,keV is ignored, adding $bbodyrad3$ only results in $\Delta \chi^{2} = 13$ with 1252 d.o.f. Considering that the accretion materials can block some emission regions, incomplete coverage of the absorption component causes soft photons to escape, which can also produce the soft excess. So we add the model, $tbpcf$, for Model-III: $constant \times tbabs \times tbpcf \times (bbodyrad1+bbodyrad2+gaussian1+gaussian2+cutoffpl) \times edge$. The fitted residuals are shown in the green dots in Figure \ref{fig:average spec} (d). Adding this model resulted in a reduction of 339 in $\chi^{2}$ with 1334 d.o.f.. If 1$-$2\,keV is ignored, adding $tbpcf$ only results in $\Delta \chi^{2} = 12$ with 1252 d.o.f.. 

For comparison with known physical models, the relativistic reflection model, \texttt{relxilllp} \citep{2014ApJ...782...76G,2016A&A...590A..76D}, is included to address the observed reflection features in Model-IV: $constant \times tbabs \times tbpcf \times (bbodyrad1+relxilllp+cutoffpl)$. \texttt{Relxilllp} models the radiation of the accretion disk reflecting the cut-off power law spectra from a lamp post geometry. Similar to \cite{2022MNRAS.516.1601B} in analyzing the \textit{NuSTAR} observations of this source, the spin parameter, $a$, velocity of lamp-post source, $\beta$, and redshift, $z$, are all set to 0, and the height of the lamp-post source, $h$, is set to 5 (in gravitational radii). The outer disk radius, $R_{\rm out}$, is set to 400 (in gravitational radii). We bind the power-law index, $\Gamma$, and cutoff energy, $E_{\rm cut}$, of the incidence spectrum to the primary spectrum. The reflection fraction, $f_{\rm refl}$, is then fixed at -1. These 4 models are listed in Table \ref{table:4 models} and the parameters of phase-averaged spectrum is list in Table \ref{table:average parameters}.

\begin{table}
\small
\caption{Phase-averaged spectrum parameters}
\label{table:average parameters}
\medskip
\begin{center}
\begin{tabular}{l l c c c c}
\hline \hline
  &  &  Model-I  &  Model-II  &  Model-III  &  Model-IV\\
\hline
tbabs & $n_{\rm H}\ (10^{22}\ \rm cm^{-2})$ & $1.02_{-0.05}^{+0.06}$ & $1.11_{-0.02}^{+0.02}$ & $0.72_{-0.03}^{+0.03}$ & $0.66_{-0.05}^{+0.04}$\\

tbpcf & $n_{\rm H_{cov}}\ (10^{22}\ \rm cm^{-2})$ & ... & ... & $1.95_{-0.23}^{+0.25}$ & $1.56_{-0.12}^{+0.14}$\\
      & $f_{\rm cov}$ & ... & ... & $0.36_{-0.02}^{+0.03}$ & $0.47_{-0.04}^{+0.05}$\\
      
bbodyrad1  &  $kT_{1}\ (\rm keV)$ & $1.34_{-0.02}^{+0.02}$ & $1.34_{-0.02}^{+0.02}$ & $1.32_{-0.03}^{+0.02}$ & $1.51_{-0.01}^{+0.01}$\\
           &  $N_{\rm bb_{1}}$ & $1236_{-59}^{+65}$ & $1171_{-46}^{+55}$ & $1084_{-58}^{+80}$ & $745_{-22}^{+23}$\\

bbodyrad2  &  $kT_{2}\ (\rm keV)$ & $3.51_{-0.34}^{+0.28}$ & $3.55_{-0.40}^{+0.32}$ & $3.43_{-0.94}^{+0.66}$ & ...\\ 
           &  $N_{\rm bb_{2}}$ & $3.18_{-0.90}^{+1.72}$ & $2.37_{-0.68}^{+1.53}$ & $1.17_{-0.75}^{+4.75}$ & ...\\
bbodyrad3  &  $kT_{3}\ (\rm keV)$ & ... & $0.117_{-0.003}^{+0.003}$ & ... & ...\\
           &  $N_{\rm bb_{3}}\ (10^{7})$ & ... & $5.14_{-0.80}^{+1.04}$ & ... & ...\\
gaussian1  &  $E_{\rm Fe_{2}}\ (\rm keV)$ & $6.60_{-0.05}^{+0.04}$ & $6.55_{-0.05}^{+0.04}$ & $6.49_{-0.06}^{+0.05}$ & ...\\
           &  $\sigma_{\rm Fe_{1}}\ (\rm keV)$ & $1.30_{-0.04}^{+0.04}$ & $1.33_{-0.04}^{+0.04}$ & $1.38_{-0.05}^{+0.05}$ & ...\\
           &  $N_{\rm Fe_{1}}$ & $0.79_{-0.05}^{+0.05}$ & $0.83_{-0.05}^{+0.05}$ & $0.91_{-0.06}^{+0.07}$ & ...\\

gaussian2  &  $E_{\rm Fe_{2}}\ (\rm keV)$ & $6.63_{-0.01}^{+0.01}$ & $6.63_{-0.01}^{+0.01}$ & $6.63_{-0.01}^{+0.01}$ & ...\\
           &  $\sigma_{\rm Fe_{2}}\ (\rm keV)$ & $0.13_{-0.02}^{+0.02}$ & $0.14_{-0.02}^{+0.02}$ & $0.14_{-0.02}^{+0.02}$ & ...\\
           &  $N_{\rm Fe_{2}}\ (10^{-2})$ & $4.0_{-0.5}^{+0.5}$ & $4.1_{-0.5}^{+0.5}$ & $4.3_{-0.5}^{+0.5}$ & ...\\

cutoffPL   &  $\Gamma$ & $1.43_{-0.01}^{+0.01}$ & $1.44_{-0.01}^{+0.01}$ & $1.47_{-0.01}^{+0.01}$ & $1.43_{-0.01}^{+0.01}$\\ 
           &  $E_{\rm cut}\ (\rm keV)$ & $24.8_{-0.2}^{+0.2}$ & $25.0_{-0.1}^{+0.1}$ & $25.3_{-0.1}^{+0.1}$ & $26.2_{-0.1}^{+0.1}$\\
           &  $N_{\rm cut}$ & $23.5_{-0.7}^{+0.7}$ & $24.7_{-0.4}^{+0.4}$ & $27.2_{-0.8}^{+0.9}$ & $17.4_{-0.4}^{+0.4}$\\

edge       &  $E_{\rm edge}\ (\rm keV)$ & $7.12_{-0.02}^{+0.02}$ & $7.13_{-0.02}^{+0.02}$ & $7.13_{-0.02}^{+0.02}$ & ...\\
           &  $\tau \ (10^{-2})$ & $5.4_{-0.6}^{+0.6}$ & $5.1_{-0.6}^{+0.6}$ & $4.7_{-0.6}^{+0.6}$ & ...\\
relxilllp  &  Incl$(^{\circ})$ & ... & ... & ... & $16.2_{-1.3}^{+1.6}$\\
           &  $R_{\rm in}$ & ... & ... & ... & $65.4_{-6.2}^{+9.5}$\\
           &  log $\xi$ & ... & ... & ... & $3.63_{-0.05}^{+0.01}$\\
           &  $A_{\rm Fe}$ & ... & ... & ... & $2.89_{-0.18}^{+0.17}$\\
           &  $N_{\rm refl}$ & ... & ... & ... & $0.29_{-0.01}^{+0.01}$\\
\hline
constant  &  ME  & $0.968_{-0.003}^{+0.003}$ & $0.969_{-0.003}^{+0.003}$ & $0.969_{-0.003}^{+0.003}$ & $0.976_{-0.003}^{+0.003}$\\
          &  HE  & $1.005_{-0.006}^{+0.005}$ & $1.006_{-0.005}^{+0.005}$ & $1.007_{-0.005}^{+0.005}$ & $1.032_{-0.004}^{+0.004}$\\
\hline
Fitting   &  $\chi^{2}_{\nu}\ (\rm dof)$  & 1.09(1252) & 1.10(1332) & 1.13(1332) & 1.33(1337)\\
\hline
\end{tabular}
\\
\end{center}
\end{table}

\subsection{Iron line profile}

\begin{figure}
    \centering
    \includegraphics[width=0.8\textwidth]{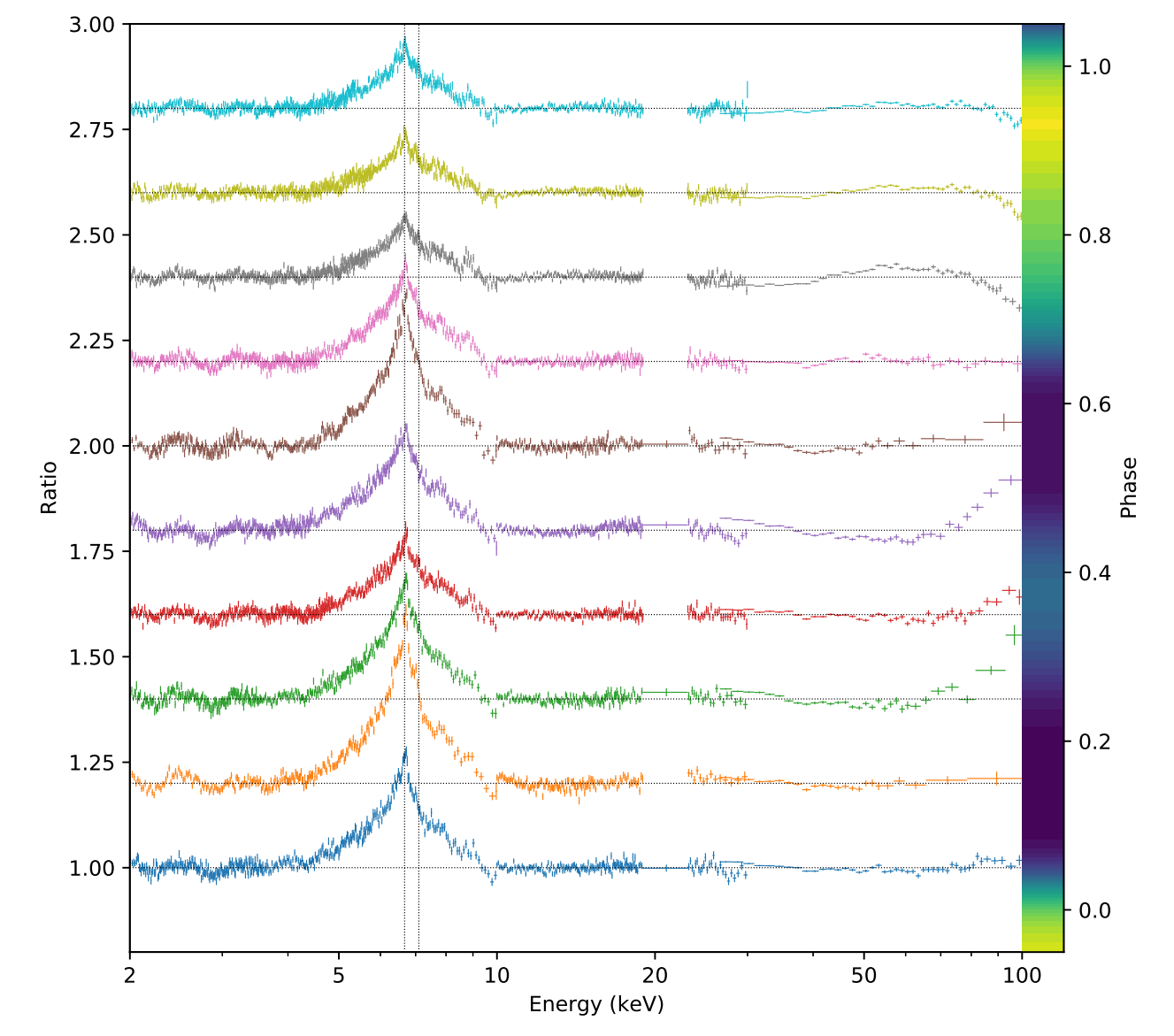}
    \caption{Ratio of observation data to the continuum model (shifts up by 0.2 in turn) in different phase intervals. The continuum is obtained by fitting 2$-$4, and 9$-$100\,keV spectrum with model: $constant \times tbabs\times(bbodyrad+cutoffpl)$. We mark the baselines corresponding to the individual ratios on the Y-axis with black dashed lines, as well as the 6.67\,keV and 7.1\,keV on the X-axis. We note that the structures shown in the residuals around 10\,keV, which is the boundary between LE and ME, likely arise from the energy range of the broad iron line not taken into account by this simple continuum modeling process. The pulse profile by HE data is plotted on the right which brighter position represents higher pulse intensity.}
    \label{fig:Fe ratio}
\end{figure}

For an overview of the phase-resolved iron line, we adopt the model: $constant \times tbabs \times (bbodyrad1+cutoffpl)$ to account for the continuum of the phase-resolved spectrum. The spectral ratios of the data with the fitted continuum model are shown in Figure \ref{fig:Fe ratio}. On the phase-resolved continuum, the hump structure in 20$-$30\,keV is obvious at phase 0.1$-$0.6. The parts of the absorption feature are present on the 80$-$100\,keV spectrum at phase 0.7$-$1.0. 

The broad iron line complex dominated by lines peaking at 6.67\,keV was reported by \cite{2019ApJ...885...18J} at the outburst peak of Swift J0243.6+6124. In their work, the iron line complex consists of a broad 6.4\,keV line, a narrow 6.67\,keV line, and a narrow 6.98\,keV line. In our fitting of the phase-averaged spectrum, the line energy of the broad iron line is approximately at 6.55\,keV (Table \ref{table:average parameters}). The iron K-edge feature at 7.1\,keV is also obvious in our ratio plot. But we did not detect the 6.98\,keV line structures in \textit{Insight}-HXMT data. As shown in these ten phase-resolved spectral ratio plots, the iron line emission is stronger at pulse-min. The equivalent width reaches 1\,keV at pulse-min but only 0.3\,keV at pulse-max. In order to quantify these pulsed properties, we still use the same models with free parameters to measure the iron lines.

\subsection{Phase-resolved spectrum}

Previously, \cite{2022ApJ...933L...3K} performed a detailed phase-resolved analysis of the cyclotron resonance scattering features (CRSF) at the peak of the outburst. We follow their fitting process but focus on the broad iron line in the lower energy range. We thus take the range of 1$-$200\,keV to fit this component with $gabs$ to obtain more accurate and appropriate continuum parameters. The width of $gabs$ is fixed at 20\,keV and thawing it has little effect on the continuum parameters. The parameters of $gabs$ are fixed when the fitting only reaches up to 100\,keV. 

\begin{figure}
    \centering
    \includegraphics[width=0.8\textwidth]{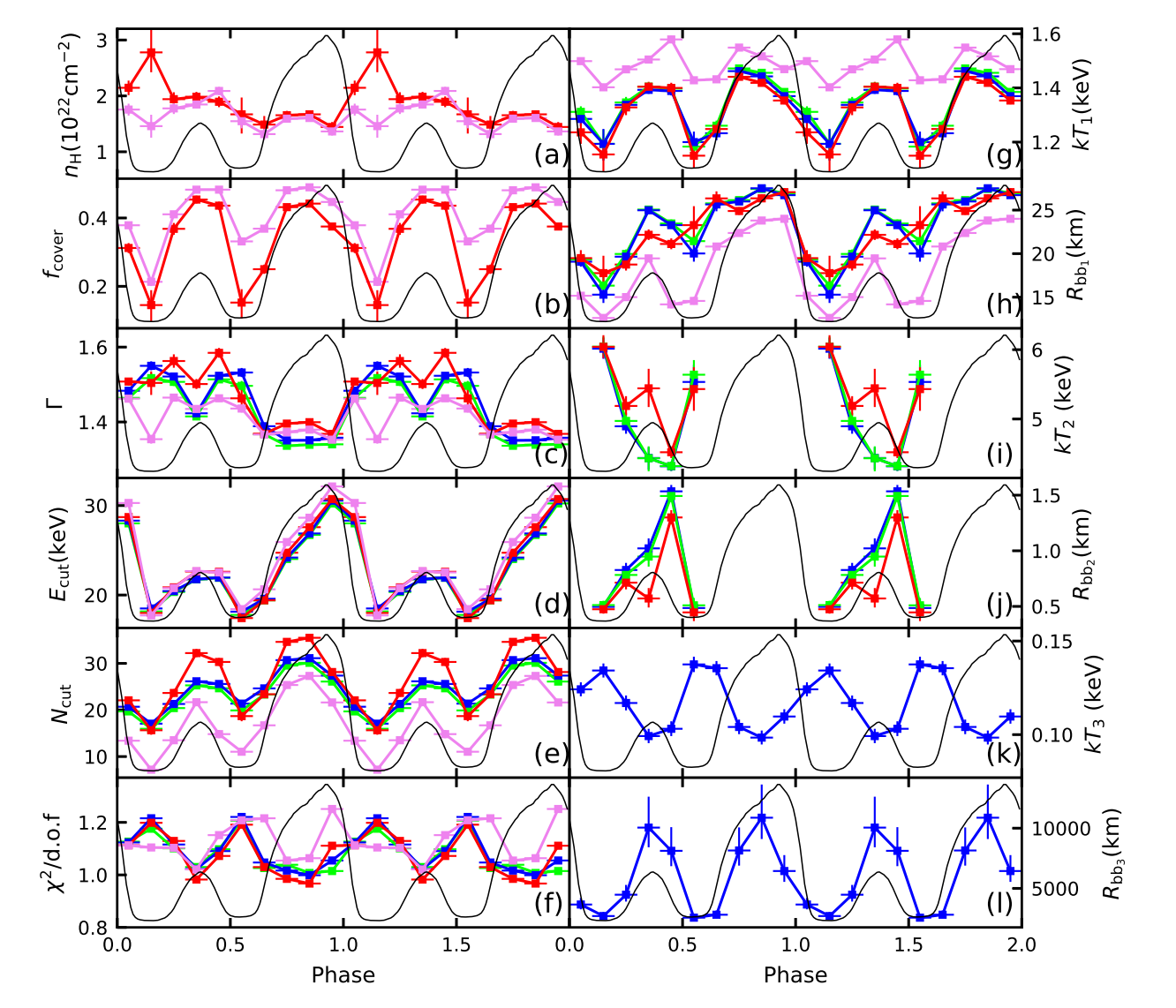}
    \caption{Comparison of continuum spectral parameters of four models. The green, blue, red, and pink data points represent models I, II, III, and IV, respectively. The pulse profile (black line) is plotted by HE data.\\
    (a)-(b) The variation of $tbpcf$ parameters in models III and IV. The column density of equivalent hydrogen, $n_{\rm H}$, in this model is positively correlated with the pulse profile. The covering factor, $f_{\rm cover}$, is also positively correlated with the pulse profile. For comparison, the fixed $n_{\rm H}$ of $tbabs$ in models I and II is drawn on panel (a).\\
    (c)-(e) The variation of $cutoffpl$ parameters. The photon index,$\Gamma$ is smaller at the main peak. Both the cutoff energy, $E_{\rm cut}$ and normalization, $N_{\rm cut}$, are positively correlated with the pulse intensity.\\
    (f) The goodness-of-fit for these four models. The error is set to 0.\\
    (g)-(l) The Blackbody parameters variation. $bb_{1}$: Both temperature ($\sim$ 1\,keV) and radius ($\sim$ 20\,km) of the blackbody are positively correlated with the pulse intensity. $bb_{2}$: The temperature $\sim$ 5\,keV, the radius is maximum in phase 0.4$-$0.5 ($\sim$ 1.5\,km), and this component is only added in model I, II, and III in phase 0.1$-$0.6. $bb_{3}$: Temperature ($\sim$ 0.1\,keV) is inversely correlated to the pulse, but the radius ($\sim$ 5000\,km) is positively correlated.
    }
    \label{fig:otherpar}
\end{figure}

The $bbodyrad2$ is only needed in the half cycle around the minor peak (phase 0.1$-$0.6). The broad hump which peaks at $\sim$ 20$-$30\,keV can be seen in phase 0.1$-$0.6 from Figure \ref{fig:Fe ratio}. It may be the Compton hump, but we use the blackbody model to fit this continuum structure. 
For spectra in other phase ranges, adding an extra blackbody component to the model improves the $\chi^{2}$ very slightly only, but has a strong coupling effect on the iron line and the determination of other continuum component. In the fitting of the phase-resolved spectra, $kT_{2}$ reaches 4$-$6\,keV in phase 0.1$-$0.6. After averaging with spectra beyond phase 0.1$-$0.6 which does not contain this component, $kT_{2}$ of the phase-averaged spectrum is only about 3.5\,keV.

\begin{figure}
    \centering
    \includegraphics[width=\textwidth]{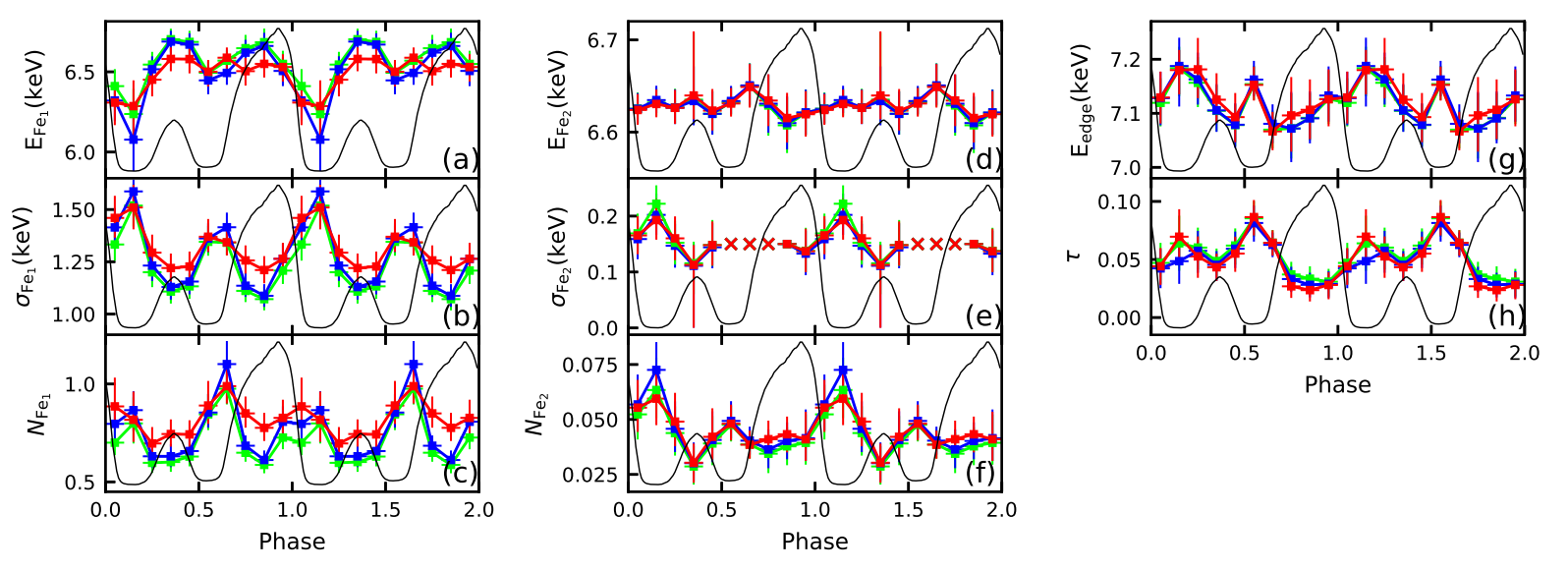}
    \caption{
    Iron line parameters versus pulsar rotation phase.\\
    The parameters obtained with Model-I , Model-II, and Model-III are in green, blue, and red respectively. For reference the black curve represents the pulse profile obtained with the HE data.
    (a)-(c) Parameters of broad iron line versus rotation phase.
    (d)-(f) Parameters of narrow iron line. $\sigma_{\rm Fe_{2}}$ is fixed at 0.15\,keV during phase 0.5$-$0.8 which are marked by cross in panel e.
    (g)-(h) Parameters of absorption edge.
    }
    \label{fig:Fe parameters}
\end{figure}

The values of $n_{\rm H}$ of the $tbabs$ for Model-I, Model-II and Model-III are fixed at 1.0, 1.1 and 0.7 (based on parameters in Table \ref{table:average parameters}), respectively, to better constrain the variation of the intrinsic spectrum. Taking into account the energy resolution, the width of $gauss2$, $\sigma_{\rm Fe_{2}}$, is difficult to constrain at phase 0.5$-$0.8 and therefore was fixed at 0.15\, keV (the cross in panel e of Figure \ref{fig:Fe parameters}.).

\begin{figure}
    \begin{minipage}[t]{0.48\textwidth}
    \begin{subfigure}
        \centering
        \includegraphics[width=\textwidth]{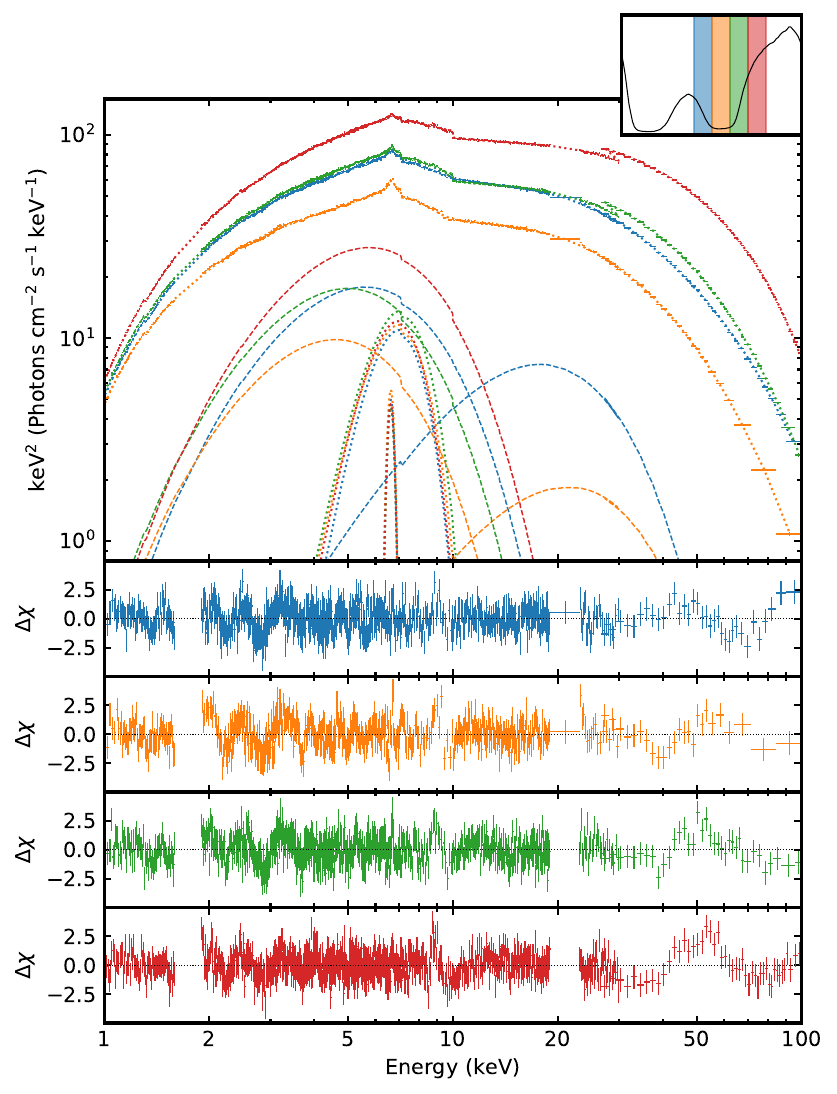}
        
    \end{subfigure}
    \end{minipage}
    \hfill
    \begin{minipage}[t]{0.48\textwidth}
    \begin{subfigure}
        \centering
        \includegraphics[width=\textwidth]{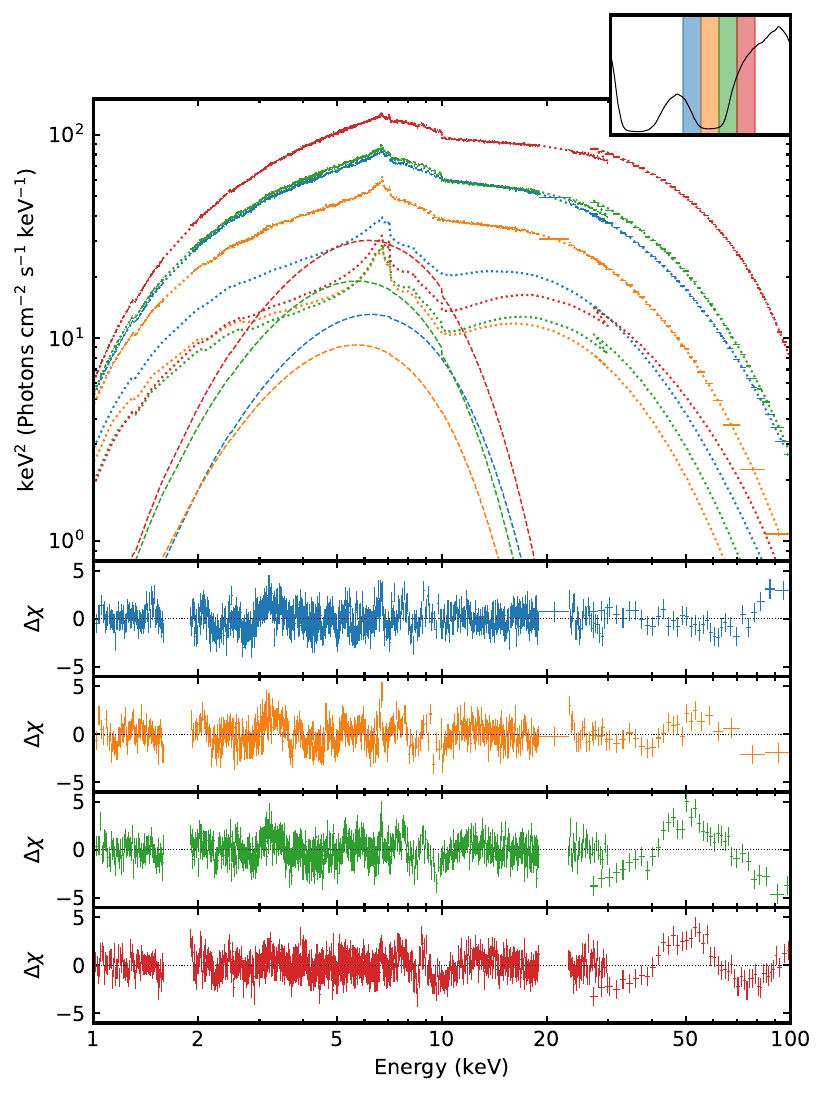}
        
    \end{subfigure}
    \end{minipage}
    \caption{The phase-resolved spectrum and reduce residuals fitted by Model-III (left panel) and Model-IV (right panel). The color of each spectra corresponds to the color of the phase interval in the pulse profile inset. Blue, orange, green, and red represent phase intervals of 0.4$-$0.5, 0.5$-$0.6, 0.6$-$0.7, 0.7$-$0.8, respectively. The dotted line draws the iron line (left panel) and reflection component (right panel). The dashed line draws the blackbody component.}
    \label{fig:spec}
\end{figure}

The continuum spectral parameters are shown in Figure \ref{fig:otherpar}.
For the fitting results of Model-II, the phase-modulated blackbody component with the inferred radius of about 2000$-$10000\,km given by Model-II seems physically implausible. For Model-III, the column density between $1.5-2.8 \times 10^{22} \,$cm$^{-2}$ corresponds to about $10^{11}$\,g variation in accretion material of the size of a neutron star. The energy range of the data is not yet able to determine this soft component, so it is only used to fit the continuum for simplicity. We only list the best-fit results for Model-III in Table \ref{table:parameter}, as the physical interpretation of this soft component is beyond the scope of this work. Some of the phase-resolved spectrum fitting parameters of interest by Model-III are shown in Figure \ref{fig:spec}. 

The best-fit iron line parameters are shown in Figure \ref{fig:Fe parameters}. Model-I, Model-II, and Model-III mainly use different components to handle the fitting in 1$-$2\,keV, but all use the spectral model consisting of two Gaussians with an absorption edge to fit for the line complex in the Fe K band. Similar results about the iron line complex are obtained, which means that our results on the iron line emission do not depend on the continuum emission models and are thus reliable. The broad line $\rm Fe_{1}$ varies periodically while the narrow line $\rm Fe_{2}$ is almost constant. Therefore, we further fixed $E_{\rm Fe_{2}}$ and $\sigma_{\rm Fe_{2}}$ to 6.63\,keV and 0.15\,keV respectively, in order to obtain a more significant variation feature of $\rm Fe_{1}$. These parameters are shown in Figure \ref{fig:sketch}. 

From the fitting parameters of the iron line, it can be seen that $E_{\rm Fe_{1}}$ varies around 6.4$-$6.6\,keV, but with a large error (Figure \ref{fig:sketch}, a). This is mainly due to the asymmetry of the iron line profile (Figure \ref{fig:Fe ratio}). The broad red wing of the iron line complex  will cause the center energy of the broad line to be lower \citep{2019ApJ...885...18J}. Therefore, the line energy and width of $\rm Fe_{1}$ appear to be anti-correlated (Figure \ref{fig:Fe parameters} and \ref{fig:sketch}). Fixing $E_{\rm Fe_{1}}$ will result in the same variation trends of iron lines, where $\sigma_{\rm Fe_{1}}$ is in the range of 1.22 to 1.42\,keV, and $N_{\rm Fe_{1}}$ is in the range of 0.67 to 1.07 (Model-III). The $\sigma_{\rm Fe_{1}}$ displays almost a half-period sinusoidal variation and is inversely correlated with the pulse intensity (Figure \ref{fig:sketch}, b). The $N_{\rm Fe_{1}}$ (count rate of the $\rm Fe_{1}$ line) curve leads the pulse profile by nearly 0.25 phase (Figure \ref{fig:sketch}, c). The absorption edge depth, $\tau$ (Figure \ref{fig:sketch}, e), presents the same variation trend with $\sigma_{\rm Fe_{1}}$ even with a fixed $n_{\rm H}$ (Model-I). We notice the unavoidable coupling between the broad iron line and the iron K-edge. But the maximum improvement in $\chi^{2}$, $\Delta\chi^{2}$, caused by adding an $edge$ component can reach 107 (with 1336 d.o.f. for Model-III) in phase 0.1$-$0.2, and lower than 20 at the peaks. Changes in the parameter of the edge optical depth, $\tau$, is responsible for this variation. It approaches to 0 in some phase intervals and cause a small value of $\Delta\chi^{2}$. In order to further study the detailed features of the iron line, we apply reflection model to provide a more physical description of the line profile and continuum.

\subsection{Reflection modeling of the phase-resolved spectra}

\begin{figure}
    \centering
    \includegraphics[width=0.4\textwidth]{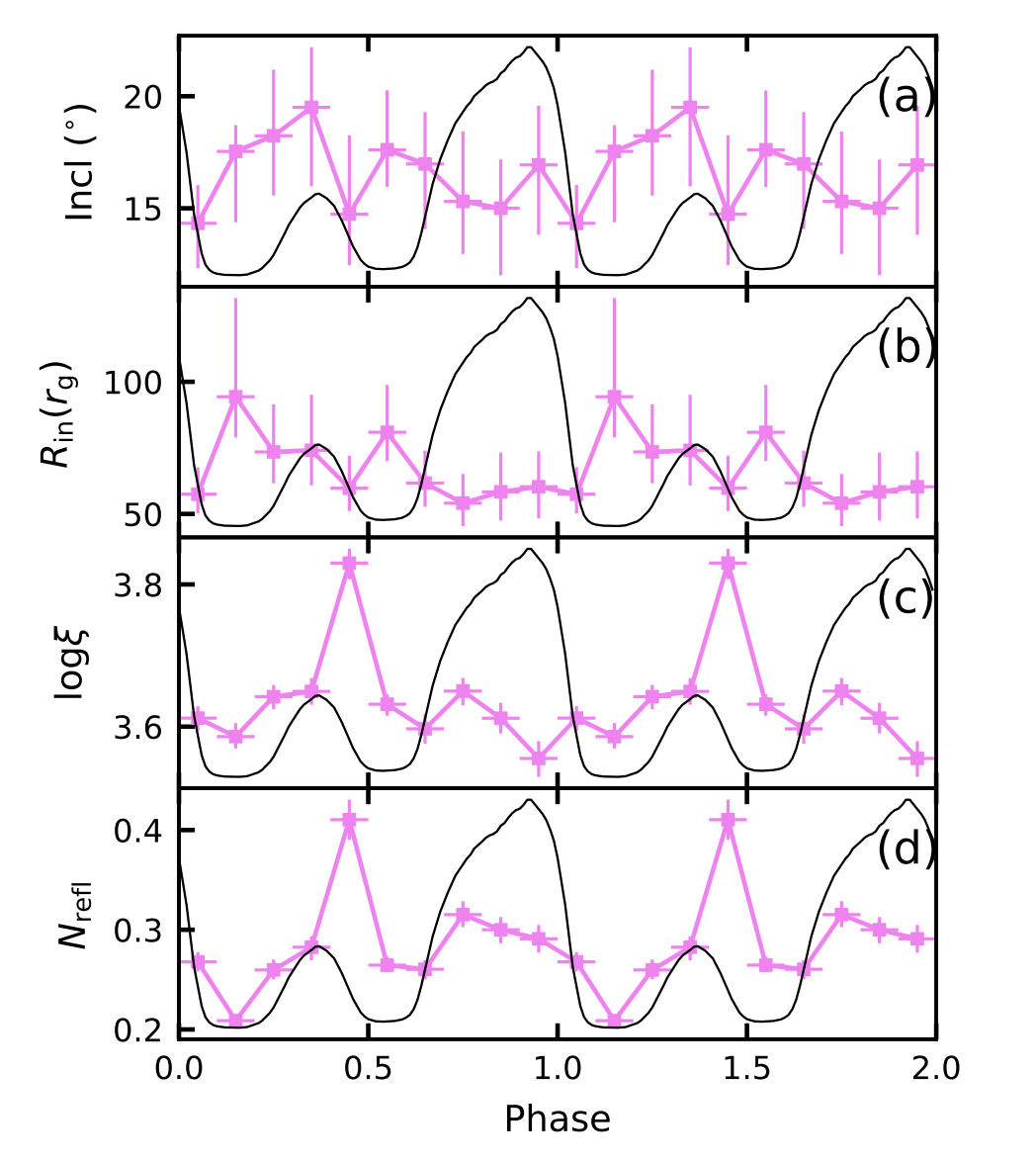}
    \caption{Parameters of reflection model versus the rotation phase. (a) Inclination angle to the normal of the disk. (b) Inner radius of the accretion disk in gravitational radii. (c) Ionization of the disk in logarithm. The value of 0 for neutral, while 4.7 for heavily ionized. (d) Normalization of the reflection component.\\
    The pulse profile (black line) is plotted by HE data. In panels (c) and (d), the parameters in phase 0.4$-$0.5 have a large shift from the mean value.}
    \label{fig:relxilllp parameters}
\end{figure}

The Iron Abundance (in Solar Units) in Model-IV is fixed at 3 to fit the phase-resolved spectra. Phase-resolved parameters of the reflection component in Model-IV are shown in Figure \ref{fig:relxilllp parameters}, and the other continuum parameters are shown in Figure \ref{fig:otherpar}. The complex line profile modeled by the reflection of the accretion disk is a good approximation of the spectrum. There may be additional reprocessing of the primary radiation for \texttt{relxilllp}, which is especially evident in phase range 0.4$-$0.5. The $bbodyrad2$ component used here has a maximum flux (Figure \ref{fig:otherpar}, panel i and j), and the ionization parameter $\log \xi$ is also peaks in this phase range. The parameter $\log \xi$ reached 3.6 (Figure \ref{fig:relxilllp parameters}, panel c) at such a high state. The inner disk radius $R_{\rm in} \sim 70\, r_{\rm g}$ (Figure \ref{fig:relxilllp parameters}, panel b) is consistent with that measured based on \textit{NuSTAR} observations \citep{2022MNRAS.516.1601B}. The gravitational radius $r_{\rm g} = 2GM/c^2$, and $r_{\rm g} = 4.13\,$km for a neutron star with $M = 1.4\,\rm M_{\sun}$. The application of reflection model of lamppost geometry provides a plausible explanation for the origin of iron line on accretion disks. In the following we will discuss the possible origin of the iron lines by combining the quantitative estimation from the $gauss$ model and the spectral characteristics of the reflection model.

\section{Discussion} \label{discuss}

Swift J0243.6+6124 reached a luminosity of $\sim10\, L_{\rm Edd}$ ($L_{\rm Edd}=1.8\times10^{38}\, \rm erg\, s^{-1}$ for a neutron star with $M = 1.4\,\rm M_{\sun}$) at its 2017$-$2018 outburst peak. In this work, we performed a phase-resolved spectral analysis during this brightest period from broad-band \textit{Insight}-HXMT observations, with the purpose of highlighting the physical properties of this PULX during the extremely high accreting states. The results show a pulsed broad iron line feature whose properties are not sensitive to the continuum spectral models. In the following we use the iron line emission features to constrain the accretion geometry of this PULX, under the framework of the radiation beam \citep{2020PASJ...72...12I} from accretion column and compact relativistic thick disk at the ultra-luminous state \citep{2020ApJ...902...18K,2020MNRAS.491.1857D,2022MNRAS.516.1601B}.

\subsection{The broad iron line profile}

The detection of iron lines by \textit{Insight}-HXMT shows that it consists of a broad line centered at around 6.5\,keV and a narrow line centered at around 6.63\,keV (consistent with \citeauthor{2019ApJ...885...18J}, but the 6.98\,keV line was not detected by \textit{Insight}-HXMT). Based on the measurements of relativistic reflection model with \textit{Insight}-HXMT data, the iron lines may originate from the accretion disk with $R_{\rm in} \sim 70\,r_{\rm g}$ and $\log\,\xi \sim 3.63$. From this perspective, the broad line could be generated in the dense inner disk region, and the narrow line should be farther away.

The 7.1\,keV absorption edge might indicate the presence of neutral iron in optically thin material around the pulsar. It is possible that the requirement of the $edge$ model component in the spectral modeling procedure arises from using the over-simplified $gauss$ model to describe the broad iron line profiles produced at the vicinity of the compact object, which usually have a slow rise and fast decay shape. A detailed discussion regarding the $edge$ model parameters would therefore be an over-interpretation. 

The inner radius of 70\,$r_{\rm g}$ is consistent with the results obtained by \textit{NuSTAR} \citep{2022MNRAS.516.1601B}. Assuming that the inner disk radius $R_{\rm in}$ is approximately the radius of the magnetosphere, the estimated dipole magnetic field is $B=1 \times 10^{12}\, \rm G$ with a coupling parameter $\Lambda=0.5$ \citep{2017MNRAS.467.1202M}. The estimated magnetic field is an order of magnitude lower than $B\sim 1.6 \times 10^{13}$\,G reported by \cite{2022ApJ...933L...3K} for the same observations, but with the detected cyclotron resonant scattering feature at around 146 keV. These two magnetic strengths may correspond to dipole and multipole fields, respectively, as also discussed by \cite{2022ApJ...933L...3K}. However, the accretion disk is most likely truncated by the magnetosphere dominated by a dipole field away from the surface of the neutron star. On the other hand, the detected cyclotron resonant scattering feature should be produced very close to the surface of the neutron star, where the multipole field dominates.

The result of Model-IV is based on the reflection geometry of the lamp post. We chose to use this model because it includes proper treatments of the general relativity effects and the atomic physics relevant to the generation of broad iron lines in accretion disks. However, this model assumes that the accretion disk is symmetrically illuminated by a lamp post source which is probably not the suitable description for the geometry of the accretion columns in neutron stars. For fan beam radiation that is unevenly distributed with phase, the variations of the direct radiation and the reflected radiation are quite probably out of sync, or even inversely correlated in extreme cases. Deviations of the illuminating source from model assumptions used as inputs in the reflection model may also lead to biased continuum parameter measurements. The occlusion of matter in the co-rotating magnetosphere (accretion curtain) is also not considered in the reflection model. Therefore, we further discuss the origin of the iron line from the periodic modulation of the Gaussian line, so as to avoid the limitation and bias of the Model-IV in describing the different phases of pulsars.

\subsection{Modulated iron line feature}

\begin{figure}
    \begin{minipage}[t]{0.48\textwidth}
    \begin{subfigure}
        \centering
        \includegraphics[width=\textwidth]{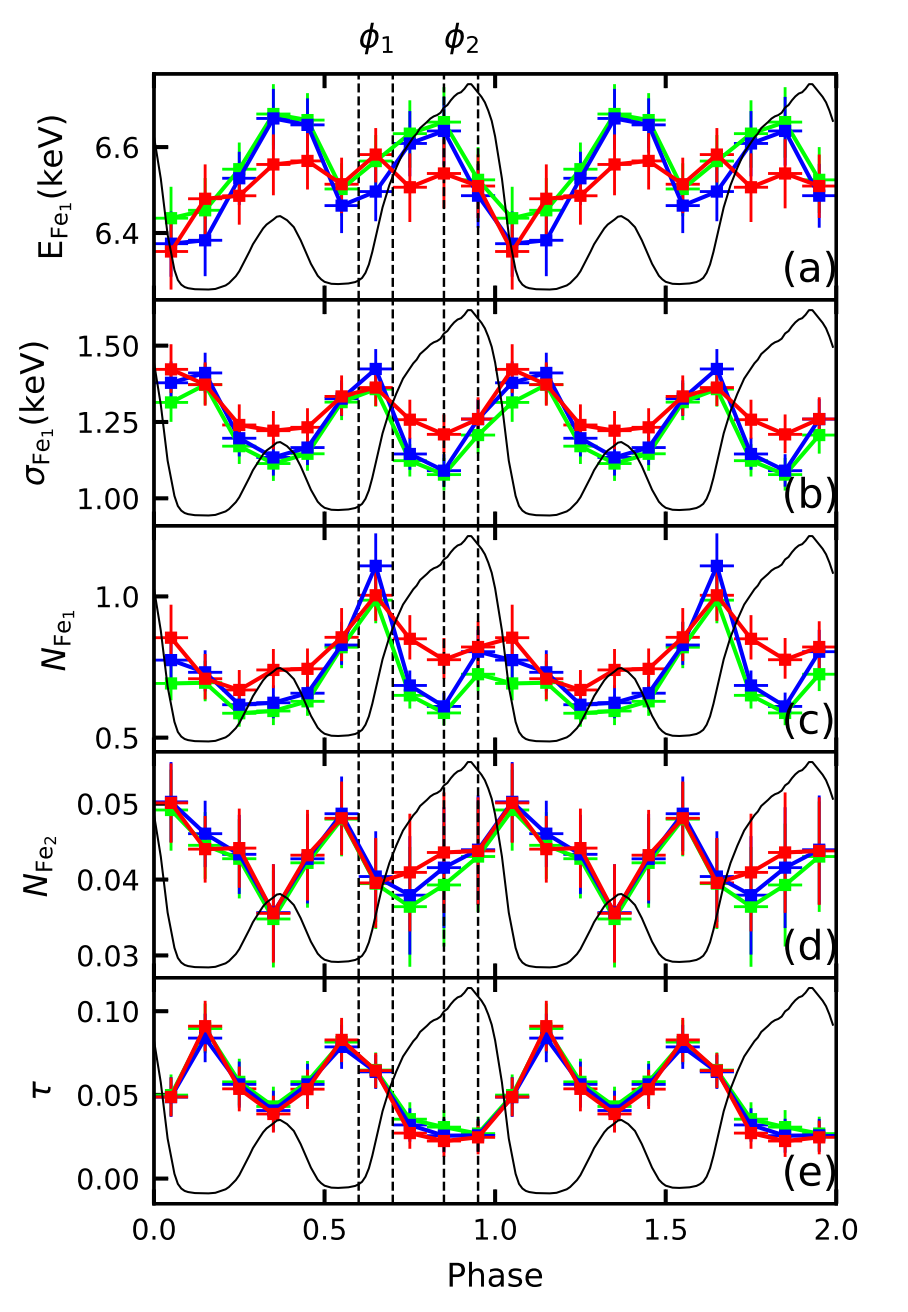}
        
    \end{subfigure}
    \end{minipage}
    \hfill
    \begin{minipage}[t]{0.48\textwidth}
    \begin{subfigure}
        \centering
        \includegraphics[width=0.95\textwidth]{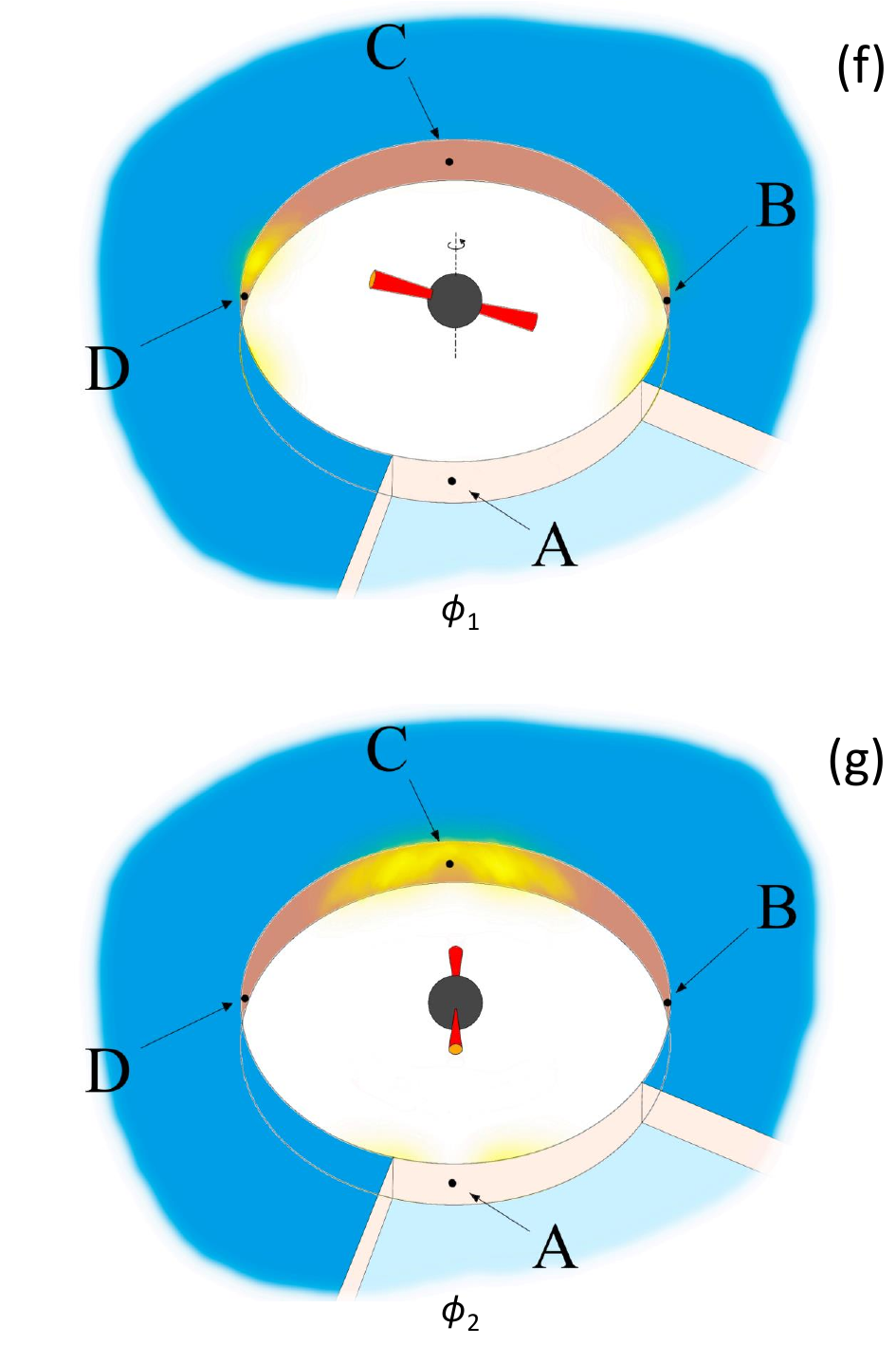}
        
    \end{subfigure}
    \end{minipage}

    \caption{Geometric sketch of the broad iron line emission region on the accretion disk of two specific phase intervals from the observer's perspective. The accretion disk is divided equally into region A, B, C, and D toroidally, and let the region A be on the side close to the line of sight of the observer.\\ 
    (a)-(e) The parameters of iron line. The results is fitting with fixed $E_{\rm Fe_{2}}$ and $\sigma_{\rm Fe_{2}}$ compared to the parameters in Figure \ref{fig:Fe parameters}. We mark two phase intervals $\phi_{1}$ (phase 0.6$-$0.7) and $\phi_{2}$ (phase 0.85$-$0.95) by the dashed boxes whose radiation configurations are illustrated in panels (f) and (g).\\
    (f) The radiation configuration in $\phi_{1}$. The radiation from accretion column is mainly concentrated in regions B and D. The apparent velocity is greater in these two regions that produce broader iron lines.\\ 
    (g) The radiation configuration in $\phi_{2}$. The radiation from accretion column is mainly concentrated in regions A and C. The apparent velocity is lower in these two regions that make a narrower line width. When considering the thickness of the disk, the fluorescent lines at the inner edge of region A will be obscured, resulting in a weaker iron line.}
    \label{fig:sketch}
\end{figure}

To discuss the possible radiation distribution of the accretion disk, we use the beam pattern of the polar cone region described by \cite{2020PASJ...72...12I}. When the angle $\theta_{\rm R} = 80^\circ$ (angle of the magnetic axis to the rotational axis) and the inclination of the rotation axis $\rm i = 50^\circ$, the simulated pulse profile (Figure 12, b in the work of \citeauthor{2020PASJ...72...12I} 2020) is similar to the structure of main and minor peaks of Swift J0243.6+6124. The radiation of fan beam directed towards the neutron star due to the electron scattering effect and gravitational bending \citep{2020PASJ...72...12I,2023hxga.book..138M}. Therefore we only draw the main radiation distribution of the fan beam (the yellow light on the inner disk in Figure \ref{fig:sketch}, f and g). Assuming that the accretion column formed at the north and south poles is symmetric with respect to the neutron star, the accretion column illuminates the disk mainly around the same azimuth direction.

Whether the iron line is modulated by the rotation of the pulsar is an important basis to determine its origin. The ratio of the iron line to the continuum (shown in Figure \ref{fig:Fe ratio}) is inversely correlated to the pulse intensity of the continuum emission. This means that the modulation amplitude of the iron line is weaker than the continuum. The iron lines should originate in a wider region than the radiation beam, such as accretion disk. The model proposed by \cite{2022MNRAS.516.1601B} provides an explanation of the reflection feature in the accretion disk of this source.

Now we discuss our results on the fluorescent iron line within the framework of this model from observer perspective, as shown in Figure \ref{fig:sketch}. Radially, the broad line is generated mainly by the gravitational red-shift of the ionized line (rather than the neutral line) from the inner disk, while the narrow line is from a more outlying region of the disk. Toroidally, the fluorescent emission of the outer disk is almost constant in different phases, resulting in a non-pulsed narrow line (Figure \ref{fig:Fe parameters}, c). The thick inner discs can be divide into 4 regions in 90$^{\circ}$ steps. We named them regions A, B, C, and D and placed region A on the side close to the line of sight of the observer as shown in Figure \ref{fig:sketch}. The regions B and D generate the mainly broadened iron line due to the stronger Doppler effect. Therefore, for the observer, the iron line properties corresponding to the geometry in $\phi_{1}$ (phase 0.6$-$0.7) is broader than in $\phi_{2}$ (phase 0.85$-$0.95), as shown in Figure \ref{fig:sketch}. If the accretion column is in $\phi_{1} + 0.5$ or $\phi_{2} + 0.5$, it will irradiate regions symmetrical with those when the accretion column is in $\phi_{1}$ or $\phi_{2}$, resulting in the same width of the iron line. This gives rise to bimodal variation of $\sigma_{\rm Fe_{\rm 1}}$. When we consider the thickness of the disk, the fluorescence surface is no longer exposed to the observer uniformly. The inner edge of region A in Figure \ref{fig:sketch} is obscured by the accretion disk itself, making the iron line in $\phi_{2}$ weaker than in $\phi_{1}$. However, parameter $N_{\rm Fe_{1}}$ in $\phi_{1}$ has a large deviation from symmetrical configuration in $\phi_{1}+0.5$. This indicates the presence of additional complex accretion structures.

The pulsed blackbody component with a temperature $kT$ of 1\,keV may correspond to an optically thick structure in the magnetosphere such as the accretion curtain or the column top \citep{2017MNRAS.467.1202M,2019ApJ...873...19T}. This structure could weaken the observed iron line by blocking the direct radiation from the column to the disk and/or fluorescence emission from the inner disk. We notice that in Figure \ref{fig:otherpar} (h) the area of $bb_{1}$ in $\phi_{1}$ is quite different from $\phi_{1}+0.5$, and thus suspect that the observed blackbody component is mainly around region A (in Figure \ref{fig:sketch}), while the blocking occurs mainly near region C. This might be the reason that the iron line in $\phi_{1}$ is stronger than that in phase $\phi_{1}+0.5$. The occlusion of the accretion column top or accretion curtain could also modulate the iron line intensity, but the distribution pattern of material on the magnetosphere is unclear, and so we did not draw this structure in Figure \ref{fig:sketch}.

We suggest that the variation of the Gaussian component is mainly caused by the inhomogeneous illumination and reflection from the thick inner disk as the pulsar rotates. The ionization state should be the same at a certain disk radius. Thus we believe the variation of $E_{\rm Fe_{1}}$ is most likely a result of the skewed line profile, with the strength of the red wing relative to the blue wing varying with phases \citep{2019ApJ...885...18J}. This may also lead to the pulsed iron K-edge, which is significant as shown by the reduction of $\chi^{2}$ (138) when adding an $edge$ model in the spectral fitting. Although the similar pulse variations were detected in V 0332+53 \citep{2021MNRAS.506.2156B}, also a Be/X-ray transient pulsar, the accretion curtain origin in their work seems out of place here. The reprocessing of optically thick envelope is difficult to form the edge structure, and in such a high state, it is difficult to observe the neutral iron atoms in this interior region. Using two Gaussian components and absorption edge can only provide a very rough and most basic estimate of the changes in the line profile. In this paper, we present clearly the phased-resolved variation behavior of the broad iron line in Swift J0243.6+6124 with high S/N spectra, and propose a basic picture to explain the origin of the line variations. However, we will not delve into the detailed physical model in this paper and leave it to future theoretical works.

\section{Summary} \label{summary}

We have performed phase-resolved spectral analysis of the ultra-luminous X-ray source Swift J0243.6+6124 in its brightest state with the data observed by \textit{Insight}-HXMT during the 2017$-$2018 outburst. In this state, it is characterized by a broad iron line peaked at $\sim$ 6.67\,keV. This profile can be approximated by a narrow Gaussian line centered at around 6.63\,keV and a broad Gaussian line centered at around 6.5\,keV. Referring to the reflection model, we speculate that the broad line originates from inner disk region ($\sim$ 70\,$r_{\rm g}$), while the narrow line originates from the outer region.

For the first time, we detect the pulsed broad iron line signature in a PULX. The variation in width and intensity of this iron line has a phase offset of about 0.25 from the pulse profile. The variation of narrow line is not significant which supports that it is from the outer disk with almost no modulation. This result is generally consistent with the radiation pattern \citep{2020PASJ...72...12I} of the dipole field and the thick disk geometry \citep{2020ApJ...902...18K,2020MNRAS.491.1857D,2022MNRAS.516.1601B}. The modulation in the broadening of the iron line is caused by the Doppler effect on the inner disk that is irradiated unevenly in different phases. Modulation in strength of broad iron line may be caused by the occlusion of the thick inner disk itself. We also propose a possibility for modulation caused by occlusion of magnetospheric material. For a more specific speculation on the modulation of the broad iron line and the results of the pulsed iron K-edge, a more detailed model description is still needed.

\begin{acknowledgments}

This work used data from the \textit{Insight}-HXMT mission, a project funded by China National Space Administration (CNSA) and the Chinese Academy of
Sciences (CAS). This work is supported by the National Key R\&D Program of China (2021YFA0718500) and the National Natural Science Foundation of China under grants U2038102, 12373051 and 12333007. This work is also supported by International Partnership Program of Chinese Academy of Sciences (Grant No.113111KYSB20190020). L. D. Kong is grateful for the financial support provided by the Sino-German (CSC-DAAD) Postdoc Scholarship Program (57251553).

\end{acknowledgments}

\bibliography{main}{}
\bibliographystyle{aasjournal}

\begin{sidewaystable}
\small
\caption{Spectral Fitting Parameters with Model-III}
\label{table:parameter}
\medskip
\begin{center}
\resizebox{\textwidth}{0.2\textwidth}{
\begin{tabular}{l l l l l l l l l l l l}
\hline \hline
Phase &  & 0.0 $-$ 0.1 & 0.1 $-$ 0.2 & 0.2 $-$ 0.3 & 0.3 $-$ 0.4 & 0.4 $-$ 0.5 & 0.5 $-$ 0.6 & 0.6 $-$ 0.7 & 0.7 $-$ 0.8 & 0.8 $-$ 0.9 & 0.9 $-$ 1.0\\
\hline
tbabs     
          & $n_{\rm H}\ (10^{22}\ \rm cm^{-2})$ & 0.7 (fixed) & 0.7 (fixed) & 0.7 (fixed) & 0.7 (fixed) & 0.7 (fixed) & 0.7 (fixed) & 0.7 (fixed) & 0.7 (fixed) & 0.7 (fixed) & 0.7 (fixed)\\
tbpcf     
          
          & $n_{\rm H_{cov}}\ (10^{22}\ \rm cm^{-2})$ & $2.14_{-0.13}^{+0.13}$ & $2.78_{-0.36}^{+0.42}$ & $1.94_{-0.13}^{+0.12}$ & $1.98_{-0.09}^{+0.09}$ & $1.89_{-0.10}^{+0.10}$ & $1.67_{-0.34}^{+0.30}$ & $1.49_{-0.16}^{+0.15}$ & $1.65_{-0.08}^{+0.08}$ & $1.67_{-0.07}^{+0.07}$ & $1.44_{-0.09}^{+0.09}$ \\

          & $f_{\rm cov}$ & $0.31_{-0.01}^{+0.01}$ & $0.15_{-0.05}^{+0.04}$ & $0.37_{-0.02}^{+0.02}$ & $0.45_{-0.01}^{+0.01}$ & $0.44_{-0.01}^{+0.01}$ & $0.15_{-0.05}^{+0.04}$ & $0.25_{-0.01}^{+0.01}$ & $0.43_{-0.01}^{+0.01}$ & $0.44_{-0.01}^{+0.01}$ & $0.37_{-0.01}^{+0.01}$ \\

bbodyrad1 
          
          & $kT_{1}\ (\rm keV)$  & $1.24_{-0.04}^{+0.03}$ & $1.15_{-0.06}^{+0.06}$ & $1.33_{-0.03}^{+0.03}$ & $1.40_{-0.02}^{+0.02}$ & $1.40_{-0.02}^{+0.02}$ & $1.15_{-0.05}^{+0.04}$ & $1.25_{-0.02}^{+0.02}$ & $1.44_{-0.01}^{+0.01}$ & $1.42_{-0.01}^{+0.01}$ & $1.35_{-0.01}^{+0.01}$ \\

          & $N_{\rm bb_{1}}$ & $821_{-64}^{+77}$ & $682_{-113}^{+153}$ & $760_{-56}^{+61}$ & $1062_{-57}^{+59}$ & $964_{-55}^{+58}$ & $1171_{-159}^{+219}$ & $1501_{-81}^{+92}$ & $1340_{-47}^{+47}$ & $1502_{-49}^{+49}$ & $1584_{-53}^{+55}$ \\

bbodyrad2 
          & $kT_{2}\ (\rm keV)$ & ... & $6.04_{-0.16}^{+0.17}$ & $5.19_{-0.14}^{+0.14}$ & $5.44_{-0.27}^{+0.28}$ & $4.52_{-0.08}^{+0.08}$ & $5.43_{-0.31}^{+0.31}$ & ... & ... & ... & ...\\
          & $N_{\rm bb_{2}}$ & ... & $0.5_{-0.1}^{+0.1}$ & $1.1_{-0.2}^{+0.2}$ & $0.7_{-0.2}^{+0.2}$ & $3.7_{-0.4}^{+0.4}$ & $0.4_{-0.2}^{+0.2}$ & ... & ... & ... & ...\\
gaussian1 
          & $E_{\rm Fe_{1}}\ (\rm keV)$ & $6.36_{-0.09}^{+0.08}$ & $6.48_{-0.09}^{+0.08}$ & $6.49_{-0.07}^{+0.07}$ & $6.56_{-0.07}^{+0.07}$ & $6.57_{-0.07}^{+0.06}$ & $6.51_{-0.07}^{+0.06}$ & $6.58_{-0.06}^{+0.06}$ & $6.51_{-0.08}^{+0.08}$ & $6.54_{-0.08}^{+0.08}$ & $6.51_{-0.07}^{+0.07}$ \\
          
          & $\sigma_{\rm Fe_{1}}\ (\rm keV)$ & $1.42_{-0.08}^{+0.08}$ & $1.37_{-0.07}^{+0.07}$ & $1.24_{-0.06}^{+0.07}$ & $1.22_{-0.06}^{+0.06}$ & $1.23_{-0.06}^{+0.06}$ & $1.33_{-0.06}^{+0.07}$ & $1.36_{-0.06}^{+0.06}$ & $1.26_{-0.06}^{+0.07}$ & $1.21_{-0.06}^{+0.07}$ & $1.26_{-0.07}^{+0.07}$ \\
          
          & $N_{\rm Fe_{1}}$ & $0.85_{-0.09}^{+0.12}$ & $0.71_{-0.07}^{+0.09}$ & $0.67_{-0.06}^{+0.07}$ & $0.74_{-0.06}^{+0.07}$ & $0.74_{-0.06}^{+0.07}$ & $0.86_{-0.08}^{+0.10}$ & $1.00_{-0.09}^{+0.11}$ & $0.85_{-0.07}^{+0.08}$ & $0.78_{-0.07}^{+0.08}$ & $0.82_{-0.08}^{+0.09}$ \\

gaussian2 
          & $E_{\rm Fe_{2}}\ (\rm keV)$ & 6.63(fixed) & 6.63(fixed) & 6.63(fixed) & 6.63(fixed) & 6.63(fixed) & 6.63(fixed) & 6.63(fixed) & 6.63(fixed) & 6.63(fixed) & 6.63(fixed)\\
          & $\sigma_{\rm Fe_{2}}\ (\rm keV)$ & 0.15(fixed) & 0.15(fixed) & 0.15(fixed) & 0.15(fixed) & 0.15(fixed) & 0.15(fixed) & 0.15(fixed) & 0.15(fixed) & 0.15(fixed) & 0.15(fixed)\\
          
          & $N_{\rm Fe_{2}}\ (10^{-2})$ & $5.01_{-0.53}^{+0.52}$ & $4.40_{-0.44}^{+0.44}$ & $4.41_{-0.52}^{+0.52}$ & $3.55_{-0.65}^{+0.65}$ & $4.32_{-0.60}^{+0.60}$ & $4.81_{-0.49}^{+0.48}$ & $3.96_{-0.59}^{+0.59}$ & $4.09_{-0.78}^{+0.77}$ & $4.36_{-0.80}^{+0.79}$ & $4.38_{-0.71}^{+0.71}$ \\

cutoffPL  
          & $\Gamma$ & $1.51_{-0.01}^{+0.01}$ & $1.50_{-0.03}^{+0.03}$ & $1.56_{-0.02}^{+0.02}$ & $1.50_{-0.01}^{+0.01}$ & $1.58_{-0.01}^{+0.01}$ & $1.46_{-0.03}^{+0.03}$ & $1.37_{-0.01}^{+0.01}$ & $1.40_{-0.01}^{+0.01}$ & $1.40_{-0.01}^{+0.01}$ & $1.37_{-0.01}^{+0.01}$ \\
          
          & $E_{\rm cut}\ (\rm keV)$ & $28.7_{-0.2}^{+0.2}$ & $18.0_{-0.3}^{+0.4}$ & $20.9_{-0.3}^{+0.3}$ & $22.6_{-0.2}^{+0.2}$ & $22.6_{-0.2}^{+0.2}$ & $17.4_{-0.3}^{+0.3}$ & $19.4_{-0.1}^{+0.1}$ & $24.7_{-0.1}^{+0.1}$ & $27.6_{-0.1}^{+0.1}$ & $30.7_{-0.2}^{+0.2}$ \\

          & $N_{\rm cut}$ & $22.1_{-0.6}^{+0.6}$ & $15.6_{-1.0}^{+1.0}$ & $23.6_{-0.8}^{+0.8}$ & $32.2_{-0.9}^{+0.9}$ & $30.3_{-0.8}^{+0.8}$ & $18.7_{-1.1}^{+1.1}$ & $23.4_{-0.5}^{+0.6}$ & $34.6_{-0.7}^{+0.7}$ & $35.5_{-0.6}^{+0.7}$ & $28.1_{-0.5}^{+0.5}$ \\

edge      & $E_{\rm edge}\ (\rm keV)$ & 7.1(fixed) & 7.1(fixed) & 7.1(fixed) & 7.1(fixed) & 7.1(fixed) & 7.1(fixed) & 7.1(fixed) & 7.1(fixed) & 7.1(fixed) & 7.1(fixed)\\
          & $\tau \ (10^{-2})$ & $4.9_{-1.2}^{+1.2}$ & $9.1_{-1.5}^{+1.5}$ & $5.4_{-1.3}^{+1.3}$ & $3.9_{-1.1}^{+1.1}$ & $5.3_{-1.2}^{+1.2}$ & $8.3_{-1.3}^{+1.3}$ & $6.5_{-1.0}^{+1.0}$ & $2.7_{-0.9}^{+0.9}$ & $2.2_{-0.9}^{+0.9}$ & $2.5_{-1.0}^{+1.0}$ \\

gabs      & $E_{\rm cyc}\ (\rm keV)$ & ... & ... & ... & ... & ... & ... & ... & 118(fixed) & 130(fixed) & 138(fixed)\\
          & $\sigma_{\rm cyc}\ (\rm keV)$ & ... & ... & ... & ... & ... & ... & ... & 20(fixed) & 20(fixed) & 20(fixed)\\
          & $S_{\rm cyc}$ & ... & ... & ... & ... & ... & ... & ... & 17.6(fixed) & 17.6(fixed) & 21.3(fixed)\\
\hline          
constant  & ME & $1.017_{-0.008}^{+0.008}$ & $1.027_{-0.010}^{+0.010}$ & $0.985_{-0.008}^{+0.008}$ & $0.959_{-0.006}^{+0.006}$ & $0.988_{-0.007}^{+0.007}$ & $1.003_{-0.009}^{+0.010}$ & $0.931_{-0.007}^{+0.007}$ & $0.927_{-0.005}^{+0.005}$ & $0.963_{-0.005}^{+0.005}$ & $0.989_{-0.006}^{+0.006}$ \\
          & HE & $0.989_{-0.009}^{+0.009}$ & $0.992_{-0.012}^{+0.012}$ & $1.011_{-0.010}^{+0.011}$ & $1.004_{-0.009}^{+0.009}$ & $1.001_{-0.009}^{+0.010}$ & $0.987_{-0.012}^{+0.012}$ & $1.006_{-0.009}^{+0.009}$ & $1.010_{-0.007}^{+0.007}$ & $1.012_{-0.007}^{+0.007}$ & $1.017_{-0.007}^{+0.008}$ \\
\hline
1-100\,keV flux & Iron line & $9.84_{-0.61}^{+0.64}$ & $7.90_{-0.18}^{+0.18}$ & $7.35_{-0.18}^{+0.18}$ & $8.11_{-0.22}^{+0.22}$ & $8.28_{-0.20}^{+0.20}$ & $9.54_{-0.21}^{+0.21}$ & $10.93_{-0.25}^{+0.25}$ & $9.13_{-0.26}^{+0.26}$ & $8.39_{-0.25}^{+0.25}$ & $8.72_{-0.25}^{+0.25}$\\

($10^{-9}$\,erg\,cm$^{-2}$\,s$^{-1}$) & Broadband & $170.66_{-0.20}^{+0.20}$ & $111.97_{-0.17}^{+0.17}$ & $165.56_{-0.19}^{+0.19}$ & $251.27_{-0.24}^{+0.24}$ & $218.11_{-0.22}^{+0.23}$ & $144.85_{-0.17}^{+0.18}$ & $216.08_{-0.21}^{+0.22}$ & $322.75_{-0.27}^{+0.28}$ & $336.29_{-0.28}^{+0.28}$ & $288.33_{-0.25}^{+0.25}$\\ 
\hline
Fitting   & $\chi^{2}_{\nu}\ (\rm dof)$ & 1.11(1338) & 1.20(1336) & 1.13(1336) & 0.98(1336) & 1.07(1336) & 1.19(1336) & 1.03(1338) & 0.99(1338) & 0.97(1338) & 1.11(1338)\\
\hline
\end{tabular}}
\\
\end{center}
\end{sidewaystable}

\end{document}